\documentstyle[emulateapj]{article}
\singlespace

\def\ref{\par\noindent\hangindent 20pt}

\def\proptosima{$\; \buildrel \propto \over \sim \;$}
\def\proptosim{\lower.5ex\hbox{\proptosima}}            
\def\ltsima{$\; \buildrel < \over \sim \;$}
\def\simlt{\lower.5ex\hbox{\ltsima}}            
\def\gtsima{$\; \buildrel > \over \sim \;$}
\def\simgt{\lower.5ex\hbox{\gtsima}}            

\def\lsim{\lower.5ex\hbox{$\; \buildrel < \over \sim \;$}}
\def\gsim{\lower.5ex\hbox{$\; \buildrel > \over \sim \;$}}

\begin{document}

\title{X--ray Spectral Variability of PKS 2005-489 During the Spectacular 
November 1998 Flare}

\author{Eric S. Perlman{\footnote{Space Telescope Science Institute, 3700 San Martin Drive, Baltimore, MD  21218.}}, 
Greg Madejski{\footnote{Laboratory for High Energy Astrophysics, 
NASA/Goddard, Greenbelt, MD  20771; also with the Department of Astronomy, 
University of Maryland, College Park.}},
John T. Stocke{\footnote{Center for Astrophysics and Space Astronomy, University of Colorado, Boulder, CO 80309.}}, \& 
Travis A. Rector{\footnote{National Optical Astronomy Observatories, 950 N. Cherry Avenue, Tucson, AZ  85719.}}}

\begin{abstract}

We report on monitoring of the BL Lac object PKS 2005$-$489 by the
Rossi X--ray Timing Explorer (RXTE) in 
October-December 1998.  During these months, the source underwent a
spectacular flare; at its peak on November 10, its 2$-$10 keV flux was
$ 3.33 \times 10^{-10} {\rm ~erg ~cm^{-2} ~s^{-1}}$, over 30 times
brighter than in quiescence.  During the rising phase, the X--ray
spectrum of PKS 2005$-$489 hardened considerably, reaching $\alpha =
1.32~ (F_\nu \propto \nu^{-\alpha})$ near maximum.  During the
declining phase, the X--ray spectrum steepened rapidly, reaching
$\alpha = 1.82$, then became somewhat harder towards the end of
December ($\alpha \sim 1.6$).  While such behavior has been seen
before, the simplicity, magnitude and duration of this flare allowed
us to study it in great detail.  We argue that this flare was caused
by either the injection of particles into the jet or {\it in situ}
particle acceleration, and that the spectral steepening which followed
the flare maximum was the result of synchrotron cooling.  Contrary to
other recently observed blazar flares (e.g., Mkn 501, 3C 279, PKS
2155-304), our results do not imply a major shift in the location of
the synchrotron peak during this flare.

\end{abstract}

Subject headings:  BL Lacertae objects: individual (PKS 2005$-$489) ---
galaxies: active --- X-rays: galaxies

\section{Introduction}

BL Lacertae objects are characterized by highly variable non-thermal
emission which dominates their characteristics from the radio through
the $\gamma$--rays.  The mechanism believed to be responsible for their
broadband emission is synchrotron radiation followed by
inverse-Compton scattering at higher energies (e.g., Blandford
\& K\"onigl 1979).  Relativistic beaming of a jet viewed at very small 
angles is the most natural explanation for the extreme properties of
the class, which include violent variability (up to 1-5 magnitudes in
the optical; see Wagner \& Witzel 1995 and references therein), high
$\gamma$--ray luminosities (Mukherjee et al. 1997), featureless optical
spectrum and superluminal motion (Vermeulen \& Cohen 1994 and
references therein).  The BL Lac class is often separated into two
subclasses via the location of the peak in the synchrotron part of
their spectrum.  High-energy peaked BL Lacs (HBLs) have synchrotron
peaks in the UV/X--ray, while low-energy peaked BL Lacs (LBLs) peak at
lower energies, in the IR/optical.  The reality and/or nature of this
apparent dichotomy is a subject of active debate in the literature
(e.g., Urry \& Padovani 1995; Padovani \& Giommi 1995; Georganopoulos
\& Marscher 1998a).

PKS 2005$-$489 is a bright, $z=0.071$ BL Lac object, with a broadband
spectrum which peaks in the UV, making it an HBL-type or
intermediate object (Sambruna et al. 1995; Perlman et
al. 1996).  As the only confirmed (few) extragalactic TeV
$\gamma$--ray sources are nearby HBLs (Catanese \& Weekes 1999 and
references therein), PKS 2005$-$489 has recently become a prime target
for southern-hemisphere Cherenkov observatories.

Here we report on the X--ray spectral variability of PKS 2005$-$489
during its November 1998 flare .  We will report on multiwaveband
observations in a future paper (Perlman et al., in preparation).  All
calculations assume $H_0 = 60 {\rm ~km ~s^{-1} ~Mpc^{-1}}$ and $q_0 =
0.1$.

\section{Observations}

RXTE observations of PKS 2005$-$489 were organized around
multiwavelength campaigns in September and October 1998.  The October
campaign began 14-15 October, intending to observe the object for 10
ks per day for five days.  Due to hardware problems, observations were
suspended 16-20 October and 23-29 October.  During the week of 18-25
October, a flare alert was issued by the all-sky monitor (ASM) team
(Remillard 1998); following consultations with the current author and
the ASM team, monitoring continued through the end of 1998, with
observations of 1-10 ks duration  every $\sim$ 2-3 days.

Two instruments aboard RXTE are designed to observe the X--ray spectrum
and variability of target sources.  The PCA (Jahoda et al. 1996)
consists of 5 co-aligned, gas-filled
proportional counter X--ray detectors (called PCUs), sensitive between
2 -- 60 keV, each of which has an open area of 1300 cm$^{2}$.  The PCA
has a roughly triangular response over its nearly 1$^\circ$ field of 
view.  During all but one observation three or
four PCUs were in use{\footnote{The exception is November 6, when PCUs 1
and 2 were in use.}}.  For maximum consistency, we use PCUs 0, 1 and
2 in our data analysis. 

The HEXTE (Rothschild et al. 1998) consists of two clusters, each
having four NaI/CsI scintillation counters, sensitive over the 15 -- 250
keV range, with a total effective area of $\sim 1600$ cm$^{2}$, and
a $\sim 1^\circ$ field of view.  The source was detected up to 40 keV
with HEXTE, but significant signal was obtained only when all HEXTE
data was combined.  For this reason, we will use only the much higher
signal to noise PCA data in the ensuing discussion.

Lightcurves were extracted using the ftools SAEXTRCT and LCURVE.  The
variable PCA background was modeled with PCABACKEST, which uses
observations of X--ray blank, high latitude areas of sky (Jahoda et
al. 1996). X--ray spectra were extracted using the script REX.  Spectral fits
were done in XSPEC v. 10.0, using all three Xenon levels and response
matrices generated by the ftool PCARMF.  Fits were done over the
ranges 3.3 -- 36.5 keV (PCA layer 1) and 10.2 -- 36.5 keV (layers 2 \&
3), ignoring channels around the 4.8 keV Xe feature.  Morrison \&
McCammon (1983) cross-sections, solar abundance and $N(H) = 5 \times
10^{20} {\rm ~ cm^{-2}}$ were assumed, consistent with SAX and ROSAT
observations (Padovani et al. in preparation; Sambruna et
al. 1995).

\section{Flux and Spectral Variability}

During the last quarter of 1998, PKS 2005$-$489 underwent a
spectacular X--ray flare, comparable to the largest observed in any BL
Lac object.  Variability of a factor $\sim 8$ was detected (Figures
1a-c; the variations are larger in the harder bands), and the source
peaked at $F(2 - 10 {\rm keV}) = 3.33 \times 10^{-10} {\rm ~erg ~cm^{-2}
~s^{-1}}$ on 10 November.  A fuller appreciation of the magnitude of
the flare is obtained by comparing with two previous ROSAT
observations (extrapolated assuming $\alpha=1.9$; Sambruna et
al. 1995) and two EXOSAT observations (Sambruna et al. 1994), which
show that a more average $2-10$ keV flux for PKS 2005$-$489 is $\sim 9
\times 10^{-12} {\rm ~erg ~cm^{-2} ~s^{-1}}$.  Somewhat higher fluxes,
respectively 3.2 and 6.1 $\times 10^{-11} {\rm ~erg ~cm^{-2}
~s^{-1}}$, were recorded during 1997 observations by RXTE and SAX
(Lamer et al. in preparation; Padovani et al., in preparation).  Thus
at its peak in 1998, PKS 2005$-$489 was about 35 times as bright as in
quiescence, with a $2-10$ keV luminosity (assuming isotropic emission)
of $4.8 \times 10^{45} {\rm ~erg ~s^{-1}}$.  By comparison, the great
1997 flare of Mkn 501 (Pian et al. 1998) represented an increase in 2
-- 10 keV flux of slightly less than a factor 20 (peak to quiescence;
{\it n. b.}, during its flare Mkn 501's synchrotron spectrum peaked at
100 keV), and a peak luminosity of $1.9 \times 10^{45} {\rm ~erg
~s^{-1}}$.

These observations show that PKS 2005$-$489 was in an enhanced flux
state for at least three months.  The flare's peak is characterized by
a plateau lasting about 4 days (Fig. 1a-c), and the half-maximum region
spans $35 \pm 5$ days.  No variability is found on timescales $<1$
day; however, our sampling on any given day covers at most 10 ks.
Both the rising and declining phases show plateaus lasting 1-2 weeks
near half maximum flux.

The most striking feature one obtains from Figure 1 is that the
emission in the harder bands declines faster.  Due to the lack of
sharp features in the lightcurve and the somewhat sparse sampling,
this is difficult to quantify via cross-correlations; indeed, by
cross-correlating the 2-5 keV and 5-10 keV lightcurves, we find no
evidence of a lag.  To crudely characterize the light curve, we take
the rise time (defined as the increase of flux from 50\% to
its peak value) to be $9 \times 10^5$ s.  For the decline, the drop of
the flux to 50\% of peak for 3, 7, and 12 keV photons (corresponding
to the weighted mean energy of the photons in the 2-5, 5-10 and 10-15
keV bands) is respectively 18, 14, and 11 $\times 10^5$ s.

A power law plus Galactic absorption was an excellent fit to the data
on all but two days, October 14 and 15.  In those cases, we fit a
broken power law model to the data.  The spectral fits
are shown in Tables 1 (single power law models) and 2 (broken
power law).  An $F$-test shows that the broken power law models cited
in Table 2 are a better representation of both the October 14 and 15
data, at $>99.9\%$ significance.

Given the more rapid drop observed in the hard X-rays, it is not
surprising that the X--ray spectrum showed significant evolution
during the flare, following a ``loop'' in the spectral index - flux
plane which has been seen in several other BL Lac flares (see \S 4 for
discussion).  In Figures 1d and 2, we show (respectively) the
progression of the X--ray spectral index with time and flux.  When
RXTE observations commenced October 14, the X--ray spectrum of PKS
2005$-$489 was already flatter than in quiescence ($\alpha = 1.45$
compared to $\alpha = 1.6-1.9$), and in fact was slightly concave,
hardening by $\Delta \alpha = 0.3$ above $\sim$ 9 keV.  The spectrum
hardened slightly as PKS 2005$-$489 continued to brighten, reaching
$\alpha=1.32$ on November 6, 4 days before the peak.  The slope
remained roughly constant during the flux maximum, and then steepened
precipitously as the flare began to decline; exactly when this change
begins is difficult to pinpoint due to the lack of observations during
November 14-20.  The steepest spectrum was observed on December 11
($\alpha=1.82$), after which the spectrum hardened once again, as the
flux continued to decrease.

\section {Discussion}

Recent observations of BL Lacs have found that flares seem to be
accompanied by a general hardening of the spectrum: not only does the
flux increase at every frequency, but the peak of the synchrotron
emission (and presumably higher-energy inverse-Compton emission) moves
further to the blue, sometimes by large factors (e.g., $\times$ 100
during a large flare of Mkn 501, Pian et al. 1998; and $\times$ 30
during a somewhat smaller flare of 1ES 2344+514; Giommi, Padovani, \&
Perlman 1999a).  How this manifests in the X--ray band depends on the
shape of each source's broadband spectrum.  The most dramatic behavior
($\Delta \alpha$ up to $\sim 0.6$) is seen in HBLs, which often follow
a ``looping'' pattern in the spectral index-flux plane.  The X--ray
spectrum hardens prior to the peak, steepens in the declining phase,
and then hardens again as the flux continues to decrease.  This
pattern has been observed in several HBLs, including Mkn 421
(Takahashi et al. 1996; Takahashi, Madejski \& Kubo 1999), PKS
2155$-$304 (Sembay et al. 1993; Chiappetti et al. 1999) and Mkn 501
(Pian et al. 1998).  There are, however, exceptions: spectral
steepening with increasing flux has been seen twice in PKS 2155$-$304
(Sembay et al. 1993).

Different behavior is observed in objects that peak at lower energies.
In LBLs, the X--ray spectrum is probably dominated by inverse-Compton
radiation (Sambruna, Maraschi, \& Urry 1996; Padovani, Giommi, \&
Fiore 1997; Kubo et al. 1998), and so only modest changes tend to be
observed, though some hardening was seen during a recent flare of BL
Lac (Madejski et al. 1999).  However, objects with peaks between
$10^{14-15}$ Hz behave very differently.  S5 0716+714 (Giommi et
al. 1999b) and AO 0235+164 (Madejski et al. 1996) exhibited X-ray
spectral {\sl steepening} in flare states (see also Urry et al. 1996),
probably because the X--ray spectrum was dominated by 
inverse-Compton emission in the low state, but by the soft ``tail'' of
the synchrotron emission in the high state.

During this flare, the X--ray spectrum of PKS2005$-$489 responded
similarly to other HBLs, but  with larger changes in the declining
phase (where $\alpha = 1.5-1.8$, similar to values obtained in quiescence)
than in the rising phase (Fig. 1d, 2).  
Importantly, our data do not imply a large shift in the location of
the synchrotron peak.  This can be seen by contrasting a low-state
peak at $10^{15.5-16}$ Hz ($\sim 0.1$ keV; Sambruna et al. 1995;
Padovani et al. 1997) with X-ray spectral indices uniformly steeper
than 1, meaning that the synchrotron peak cannot have moved to $\gsim
10^{16.5-17}$ Hz, i.e., a factor 10.  This flare of PKS 2005$-$489
thus appears to have been more similar to flares of Mkn 421 (Takahashi
et al. 1999) than of Mkn 501 (Pian et al. 1998) or 1ES 2344+514
(Giommi et al. 1999a), where much larger shifts in the spectral peak
were observed.  This could indicate a diversity in physical
circumstances in blazar flares such as the balance between
acceleration, particle injection and cooling.

The usual interpretation of the X--ray spectral changes observed in
HBLs (see models by Georganopoulos \& Marscher 1998b; Kirk,
Mastichiadis, \& Rieger 1998; and Coppi \& Aharonian 1999) is that the
hardening observed during the flare's rising phase is caused by
reacceleration and/or injection of electrons (either freshly
injected or already within the jet).  This may be associated with a local
compression or augmentation of the jet magnetic field (as has recently
been observed in HST polarimetry of knots of the M87 jet which have
superluminal components; see Perlman et al. 1999).  In the declining
phase, synchrotron or Compton cooling is the dominant factor.

We therefore interpret the flare as follows: the {\sl increase} of the
flux is due to a fresh injection or acceleration of energetic
particles throughout the source, which we assume happened on a time
scale that is (locally) much shorter than the light travel time across
the source.  In this context, the time scale of the increase of the
flux $\times c$ may well be an indication of the source size.  The
presence of a somewhat harder component in the spectra of October 14
and 15 indicates considerable structure in the flaring region which
these data could not resolve.  This component may be associated with
the onset of either a second, high-energy synchrotron component, or
the onset of Comptonization.  The {\sl decrease} of the flux -- which
is slower than the rise for all energies -- is caused by those
particles cooling via synchrotron radiation.  No $\gamma$--ray
emission has so far been detected for PKS 2005$-$489, so the
synchrotron component is most likely more luminous than the Compton
component.  We therefore assume that the synchrotron losses dominate.

As we mentioned above, it is now widely accepted that the broadband
emission in blazars arises in a relativistic jet pointing close to the
line of sight.  This results in 
observed time scales being shortened, and the detected photons being 
blueshifted by a factor $\delta = [\Gamma \times (1-\beta {\rm
cos}\theta)]^{-1}$, where $\Gamma$ is the bulk Lorentz factor of the
jet, $\beta = v/c$, and $\theta$ is the jet viewing angle with respect
to the observer.  Furthermore, since the emission is not
isotropic (and relativistically boosted in the direction of motion
of the jet), the luminosity inferred under an assumption of isotropy
is an overestimate.  We have no information regarding $\delta$ for PKS
2005$-$489, but in the ensuing discussion, we take $\delta = 5$ to 10,
which is the value inferred for many other blazars (e.g., Ulrich,
Maraschi, \& Urry 1997).

For a homogeneous emitting region, the radiative lifetime of a
relativistic electron emitting synchrotron photons with energy
$E_{\rm keV}$ is (in the observer's frame) $\tau_{sync} = 1.2 \times
10^3 B^{-3/2} E_{\rm keV}^{-1/2} \delta^{-1/2}$ s (cf. Rybicki \&
Lightman 1979).  This should give us at least some estimate of the
magnetic field $B$.  Of course we do not know the extent to which the
time scale of the flux decrease was due to the propagation of the
signal throughout the source, and to what extent it was caused by the
synchrotron cooling.  However, our data allows us to measure the drop
of the flux in three energy bands, where $\tau_{1/2} (E)$ = 18, 14,
and 11 $\times 10^{5}$ s respectively for 3, 7, and 12 keV photons
(cf. Fig. 1).  Following Takahashi et al. (1996), Fig. 4, we can write
$\tau_{sync} (E) - \tau_{sync}($12 keV$) = 1.2 \times 10^{3} B^{-3/2}
\delta^{-1/2} (E_{\rm keV}^{-1/2} - 12^{-1/2})$.  Comparing the
decline of the 7 keV and 12 keV photons, we infer $B = 0.003$
$(\delta/5)^{-1/3}$ Gauss; comparing the decline of the 3 keV vs. 12
keV photons, we obtain $B=0.004 ~(\delta/5)^{-1/3}$ Gauss.  With $B = 4
\times 10^{-3}$ G, the Lorentz factors of the electrons $\gamma_{\rm
el}$ radiating at energy $E$ can be estimated from $E = 4 \times
10^{-15} \gamma_{\rm el}^{2} (\delta/5)$ keV.  This implies that
$\gamma_{\rm el}$ of electrons radiating in the X--ray band is in
excess of $10^7$.  (Note that $\gamma_{\rm el}$ refers to the Lorentz
factors of individual radiating particles as distinct from $\Gamma$.)

The value of $B$ calculated as above is significantly lower than 0.2
Gauss, and $\gamma_{\rm el}$ higher than $5 \times 10^{5}$ inferred
for Mkn 421 by Takahashi et al. (1996).  This may be either due to a
real difference between the two sources (note that the inferred
isotropic luminosity of PKS 2005$-$489 is a factor of a few greater
than that of Mkn 421), or because $\tau_{sync}$ inferred by us is an
underestimate.  This last possibility is likely, if the flare observed
by the RXTE consists of a superposition of multiple, rapid variability
(shorter than $\sim$ a day) events unresolved by this observation.  An
absence of such short-timescale variability would be rare for BL Lacs
(Wagner \& Witzel 1995).  If in fact variability on timescales shorter
than a day {\sl is} present, the value of $B$ we calculate is instead
a lower limit, and $\gamma_{\rm el}$ is an upper limit.  A conclusive
test of the physical parameters in this source would be a detection of
the TeV $\gamma$--ray emission, and any correlation with the X--ray
flux.

These uncertainties underline the importance of future campaigns and
modeling efforts.  Better, more dense sampling might tell us if a
large flare actually consists of many smaller but more rapid flares.
The differences between the behavior of PKS 2005$-$489 during this
flare, and that of other flaring blazars, point out that as yet we do
not understand well how blazar jets respond to stimuli.  Such models
will need to include a large number of factors, meaning not only the
injection and reacceleration of particles, but also compression of
magnetic fields in shock regions (which both accelerates electrons and
shortens their dynamical lifetimes).  Several authors have begun to
build models which take these effects into account (e.g., Kirk et
al. 1998, Georganopoulos \& Marscher 1998b); however intensive
multiwavelength monitoring of many more flares (in both HBL and LBL
type objects) will be needed to reach a complete understanding.

\begin{acknowledgements}

We wish to thank the RXTE ASM team at MIT, and the RXTE team at GSFC,
for their rapid response to the flare of PKS 2005$-$489 during October
1998 and for continuing observations through December 31.  We wish to
thank the staff of the RXTE GOF, and particularly Tess Jaffe, for
their help in reducing the data from these observations.
E. S. P. acknowledges support from NASA grant NAG5-7122.  We wish to
thank our referee, Rita Sambruna, for many helpful comments which
significantly improved this paper.

\end{acknowledgements}

\begin{deluxetable}{cclc}
\tablewidth{0pc}
\tablecolumns{4}
\tablecaption{Single Power Law Spectral Fits}
\tablehead{
\colhead{Date (1998)} & \colhead{$F$ (2-10 keV)}  & 
\colhead{$\alpha$} & \colhead{$\chi^2_\nu$ (133 channels)} \\
& \colhead {$(\times 10^{-11} {\rm ~erg ~cm^{-2} ~s^{-1}})$} 
& \colhead {$(F_\nu \propto \nu^{-\alpha})$}
}

\startdata

10/14&	16.510&	$1.420\pm 0.011^1$ & 1.247	\nl
10/15& 	16.776&	$1.368^{+0.011}_{-0.010}$ & 1.505	\nl
10/21& 	16.370&	$1.386 \pm 0.013$ & 0.742	\nl
10/22& 	15.402&	$1.444 \pm 0.015$ & 0.739	\nl
10/30& 	20.730&	$1.421^{+0.012}_{-0.013}$ & 0.529	\nl
10/31&	20.930&	$1.408 \pm 0.012$ & 0.699	\nl
11/01& 	21.273&	$1.374 \pm 0.010$ & 0.509	\nl
11/02& 	22.783&	$1.374 \pm 0.013$ & 0.737	\nl
11/04& 	26.687&	$1.333^{+0.018}_{-0.019}$ & 0.911	\nl
11/06& 	29.845&	$1.321^{+0.015}_{-0.014}$ & 0.559	\nl
11/08& 	32.848&	$1.350 \pm 0.014$ & 0.728	\nl
11/10& 	33.327&	$1.350 \pm 0.016$ & 0.763	\nl
11/12& 	31.910&	$1.343 \pm 0.008$ & 1.021	\nl
11/14& 	29.525&	$1.371 \pm 0.016$ & 0.600	\nl
11/20& 	25.116&	$1.476^{+0.016}_{-0.015}$ & 0.570	\nl
11/22& 	22.982&	$1.556 \pm 0.018$ & 0.631	\nl
11/23& 	21.626&	$1.573 \pm 0.019$ & 0.699	\nl
11/26& 	18.472&	$1.587 \pm 0.023$ & 0.758	\nl
11/28& 	18.449&	$1.671^{+0.016}_{-0.017}$ & 0.857	\nl
11/30& 	16.003&	$1.683^{+0.019}_{-0.020}$ & 0.610	\nl
12/06& 	13.796&	$1.767 \pm 0.023$ & 0.548	\nl
12/11& 	15.933&	$1.819^{+0.027}_{-0.026}$ & 0.715	\nl
12/13& 	14.055&	$1.702 \pm 0.028$ & 0.829	\nl
12/15& 	13.631&	$1.638^{+0.084}_{-0.082}$ & 0.424	\nl
12/19& 	10.949&	$1.645 \pm 0.077$ & 0.564	\nl
12/22& 	 7.540&	$1.594^{+0.083}_{-0.082}$ & 0.560	\nl
12/23& 	 6.703&	$1.572^{+0.061}_{-0.062}$ & 0.637	\nl
12/25& 	 5.804&	$1.647^{+0.067}_{-0.066}$ & 0.627	\nl
12/27& 	 5.013&	$1.610^{+0.076}_{-0.074}$ & 0.511	\nl
12/29& 	 4.452&	$1.596^{+0.047}_{-0.046}$ & 0.546	\nl
12/31& 	 4.252&	$1.604^{+0.063}_{-0.062}$ & 0.602	\nl

\enddata

\tablenotetext {1}  {All errors given are 90\% confidence.}

\end{deluxetable}

\begin{deluxetable}{cclllc}
\tablewidth{0pc}
\tablecolumns{6}
\tablecaption{Broken Power Law Spectral Fits}
\tablehead{
\colhead{Date (1998)} & \colhead{F (2-10 keV)} & 
\colhead{$\alpha_{soft}$} & \colhead{$\alpha_{hard}$} & \colhead{$E_{break}$} & 
\colhead{$\chi^2_\nu$ (133 channels)} \\
& \colhead{$(\times 10^{-11} {\rm ~erg ~cm^{-2} ~s^{-1}})$}}

\startdata

10/14 & 16.50 & $1.45^{+0.02}_{-0.01}$ & $1.12^{+0.10}_{-0.18}$ & $9.6^{+1.8}_{-1.1}$ & 0.840 \nl
10/15 & 16.77 & $1.40^{+0.02}_{-0.01}$ & $1.14^{+0.07}_{-0.11}$ & $8.9^{+1.3}_{-1.0}$ & 1.104 \nl

\enddata

\end{deluxetable}

\begin{figure}
\plottwo{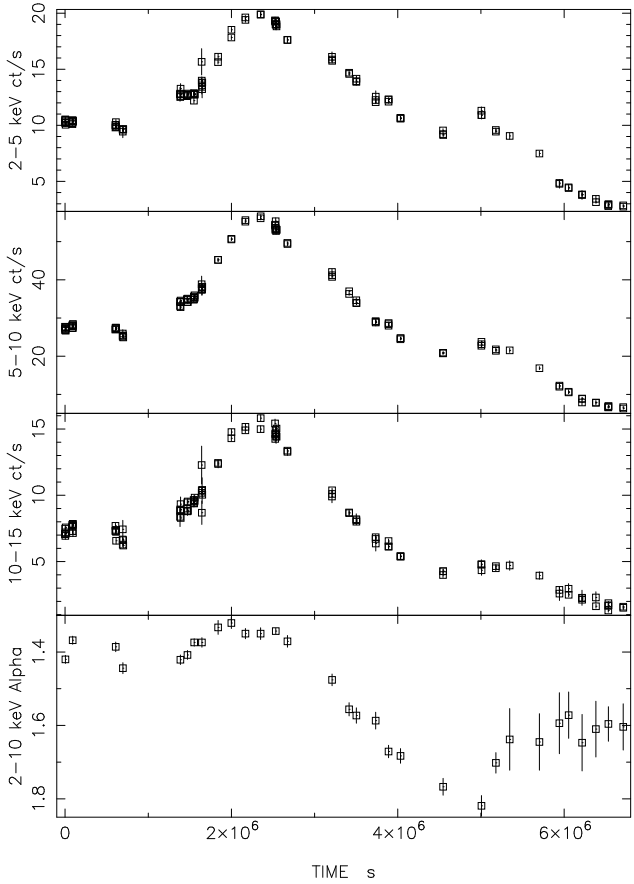}{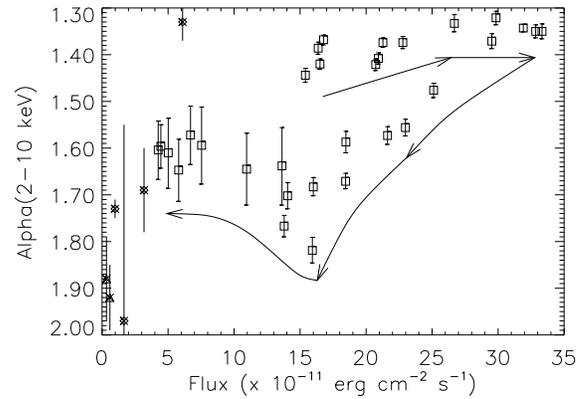}
\caption[1]{Here we show the RXTE PCA count rates and energy
spectral indices for PKS 2005$-$489 during October-December 1998.
Count rates are given in the 2-5 keV (top, Fig. 1a), 5-10 keV (second
from top, Fig. 1b) and 10-15 keV bands (third from top, Fig 1c), while
the spectral indices we plot (bottom, Fig. 1d) are for the 2-10 keV
band.  Given the observed spectra, the weighted mean energy of the
photons in the 2-5 keV band is $\sim 3$ keV, in the 5-10 keV band it
is $\sim 7$ keV, and in the 10-15 keV band it is $\sim 12$ keV.  These
values are spectral index dependent, but change only a few percent for
the spectra observed.  On this Figure, a time of zero seconds refers 
to the beginning of the first (October 14) observation, and the peak,
which occurred on November 10, is at $2.4 \times 10^6$ s.}

\caption[2]{The progression of spectral index with flux during the
1998 flare.  For comparison, we show ROSAT and EXOSAT observations
(Sambruna et al. 1994, 1995) as triangles, and previous RXTE and SAX
observations (Lamer et al. in prep; Padovani et al. 1999) as diamonds.
All error bars are 90\% confidence.  The arrows indicate the general
time sequence.  }

\end{figure}

\end{document}